\documentclass{article}
\usepackage[T1]{fontenc}
\usepackage{wrapfig}
\usepackage[portuges,english]{babel}
\usepackage{amssymb,latexsym,amsmath,color,mathrsfs,ifsym,graphics,stmaryrd} 
\usepackage[colorlinks,linkcolor=blue,urlcolor=blue,citecolor=black,
plainpages=false,pdfpagelabels,breaklinks]{hyperref}

\title{Representational Realism, Closed Theories\\and the Quantum to Classical Limit}
\author{{\sc Christian de Ronde}\thanks{Fellow Researcher of the Consejo
Nacional de Investigaciones Cient\'{\i}ficas y T\'ecnicas.}}
\date{\begin{center}
\begin{small} 
CONICET, Buenos Aires University - Argentina \\
Center Leo Apostel and Foundations of  the Exact Sciences\\
Brussels Free University - Belgium \\
\end{small}
\end{center}}
\begin{document}
\maketitle

\begin{abstract}
\noindent In this paper we discuss the representational realist stance as a pluralist ontic approach to inter-theoretic relationships. Our stance stresses the fact that physical theories require the necessary consideration of a conceptual level of discourse which determines and configures the specific field of phenomena discussed by each particular theory. We will criticize the orthodox line of research which has grounded the analysis about QM in two (Bohrian) metaphysical presuppositions ---accepted in the present as dogmas that all interpretations must follow. We will also examine how the orthodox project of ``bridging the gap'' between the quantum and the classical domains has constrained the possibilities of research, producing only a limited set of interpretational problems which only focus in the justification of ``classical reality'' and exclude the possibility of analyzing the possibilities of non-classical conceptual representations of QM. The representational realist stance introduces two new problems, namely, the superposition problem and the contextuality problem, which consider explicitly the conceptual representation of orthodox QM beyond the mere reference to mathematical structures and measurement outcomes. In the final part of the paper, we revisit, from a representational realist perspective, the quantum to classical limit and the orthodox claim that this inter-theoretic relation can be explained through the principle of decoherence. 
\medskip
\end{abstract}


\renewenvironment{enumerate}{\begin{list}{}{\rm \labelwidth 0mm
\leftmargin 0mm}} {\end{list}}

\newcommand{\ita}{\textit}
\newcommand{\mcal}{\mathcal}
\newcommand{\mfrak}{\mathfrak}
\newcommand{\mbb}{\mathbb}
\newcommand{\mrm}{\mathrm}
\newcommand{\msf}{\mathsf}
\newcommand{\mscr}{\mathscr}
\newcommand{\lra}{\leftrightarrow}
\renewenvironment{enumerate}{\begin{list}{}{\rm \labelwidth 0mm
\leftmargin 5mm}} {\end{list}}

\newtheorem{dfn}{\sc{Definition}}[section]
\newtheorem{thm}{\sc{Theorem}}[section]
\newtheorem{lem}{\sc{Lemma}}[section]
\newtheorem{cor}[thm]{\sc{Corollary}}
\newcommand{\Proof}{\textit{Proof:} \,}
\newcommand{\cqd}{{\rule{.70ex}{2ex}} \medskip}

\newpage

\section{The Classical Representation of Physical\\ Reality and QM}

The general characterization and representation of the idea of ``classical reality'', which encompases the whole of classical physics  (including relativity theory), can be condensed in the notion of {\it Actual State of Affairs (ASA)}.\footnote{See for discussion and definition of this notion in the context of classical physics \cite{RFD14}.} This particular representation was developed by physics since Newton's mechanics and can be formulated in terms of {\it systems constituted by a set of actual (definite valued) preexistent properties.} Actual classical properties are defined in terms of three logical and ontological principles proposed by Aristotle. Indeed, the Principle of Existence (PE), the Principle of Non-Contradiction (PNC) and the Principle of Identity (PI) founded not only classical logic itself but also the basis of our classical meta-physical understanding of the world. In fact, during the 18th century these principles were also used implicitly by Newton in order to provide a metaphysical definition of the notion of actual entity in classical mechanics.\footnote{For a detailed analysis of the relation between these metaphysical principles and Newton's representation of physical relaity see \cite{RFD14b}.} Through these principles Newton was able to conceive a Universe constituted by bodies which existed, possesed non-contradictory properties and remained always identical to themselves.\footnote{As noticed by Verelst and Coecke \cite[p. 167]{VerelstCoecke}, these principles are ``exemplified in the three possible usages of the verb `to be': existential, predicative, and identical. The Aristotelian syllogism always starts with the affirmation of existence: something is. The principle of contradiction then concerns the way one can speak (predicate) validly about this existing object, i.e. about the true and falsehood of its having properties, not about its being in existence. The principle of identity states that the entity is identical to itself at any moment (a=a), thus granting the stability necessary to name (identify) it.''} This physical representation of reality also allowed Laplace to imagine a demon who, given the complete and exact knowledge of all particles in the Universe (i.e., the set of all actual properties), would have immediate access, through the equations of motion, to the future and the past of the whole Universe.\footnote{Acording to the French physicist and mathematician Pierre Simon Laplace \cite{Laplace02}: ``We may regard the present state of the universe as the effect of its past and the cause of its future. An intellect which at any given moment knew all of the forces that animate nature and the mutual positions of the beings that compose it, if this intellect were vast enough to submit the data to analysis, could condense into a single formula the movement of the greatest bodies of the universe and that of the lightest atom; for such an intellect nothing could be uncertain and the future just like the past would be present before its eyes.''}

However, it is well known that in the case of QM we have serious difficulties to interpret the formalism in terms of ``classical reality''. An evidence of the deep crisis of physical representation within QM is the fact that more than one century after its creation the physics community has reached no consensus about what the theory is really talking about. A possibility with many adepts today, is to simply deny the need of conceptual representation within physics and understand the discipline as a mathematical scheme restricted to the prediction of measurement outcomes. This idea goes back to logical positivism and the Machian understanding of physical theories as an ``economy of experience'' with no metaphysical background. Taking a radical epistemic viewpoint Fuchs and Peres \cite[p. 70]{FuchsPeres00} have argued that ``[...] quantum theory does not describe physical reality. What it does is provide an algorithm for computing probabilities for the macroscopic events (`detector clicks') that are the consequences of experimental interventions. This strict definition of the scope of quantum theory is the only interpretation ever needed, whether by experimenters or theorists.'' This instrumentalist perspective is satisfied with having a empirically adequate ``algorithmic recipe''. There is no need for an interpretation because the mathematical structure already provides the correct predictions for measurement outcomes. And that is what physics is all about. 

In contraposition to that radical epistemic perspective, in this paper we discuss a representational realist stance according to which the task of both physics and philosophy of physics is to produce conceptual representations which allows us to understand and explain the features of the world and reality beyond mathematical structures and measurement outcomes. According to our stance, physical representation requires the coherent interelation between mathematical formalisms and conceptual schemes. Thus, representational realism imposes the need of introducing adequate physical concepts in order to understand QM beyond the mere reference to mathematical structures and measurement outcomes. If we accept this challenge there seems to be two main lines of research to consider. The first one is to investigate the possibility that QM makes reference to the same physical representation provided by classical physics; i.e. that it talks about ``classical reality''. This is the main idea presupposed, for example, by the hidden variables program which, as noticed by Bacciagaluppi  \cite[p. 74]{Bacciagaluppi96}, attempts to ``restore a classical way of thinking about {\it what there is}.'' The second line would be to consider the possibility that QM might describe physical reality in a different ---maybe even incommensurable--- way to that of classical physics. Apart from some few unorthodox attempts (e.g., \cite{Aerts09, Kastner12}), this second line of research has not been truly addresed within orthodoxy. The reason, we believe, is related to the introduction by Niels Bohr of two metaphysical presuppositions which assume, as a general stanpoint, the idea that the only possible representation of experience and physical reality is necessarily provided by the language of classical physics. We will analyse these strong metaphysical presuppositions in the following section. We remark that our analysis is only concerned with the philosophy of QM and does not attempt to derive from it a general conclussion regarding the general debate already present in philosophy of science.

\section{The Two Bohrian Dogmas of QM and the\\Attempt to Restore ``Classical Reality''}

In \cite{deRonde15b}, we argued that Bohr is the main responsible for producing an epistemological interpretation of QM that does not only limit physical representation in terms of classical language and classical phenomena but also precludes the very possibility of introducing and developing new (non-classical) concepts. Bohr is responsible for having introduced in the debate about the interpretation of QM two strong metaphysical presuppositions which have constrained the possibilities of analysis and development of quantum theory. Since the mid 20th century what became orthodoxy, took these ideas as necessary standpoints to think about any possible interpretation of QM. Accepted as unquestionable, these metaphysical presuppositions were turned into dogmas that all interpretations of QM would have to respect.

The first metaphysical dogma is the idea that there must exist a ``quantum to classical limit'', implying what Alisa Bokulich calls an ``open theory approach'' \cite{Bokulich04}. This reductionistic inter-theoretic relation between QM and classical mechanics was introduced by Bohr through his {\it correspondence principle}.\footnote{For a detailed analysis of Bohr's correspondence principle see \cite{BokulichCP}.}

\begin{enumerate}
{\bf \item[Dogma I.]  Quantum to Classical Limit:}  The principle that there must exist a continuous ``bridge'' or ``limit'' between classical mechanics and QM.
\end{enumerate}

\noindent However, this is certainly not the only possible way to approach the problem of inter-theory relation. In fact, this reductionistic understanding of the relation between classical mechanics and QM has been severly questioned by many pluralist acocunts in the last decades. Pluralists approaches open different possibilities of analysis which due to the single-viewed orthodox perspective have not been discussed nor developed within the literature. 

The second metaphysical dogma which has guided orthodoxy can also be traced back to Bohr's claim that physical experience needs to be expressed exclusively in terms of classical physical language \cite{BokulichBokulich}. Bohr \cite[p. 7]{WZ} stated that: ``[...] the unambiguous interpretation  of any measurement must be essentially framed in terms of classical physical theories, and we may say that in this sense the language of Newton and Maxwell will remain the language of physicists for all time.'' In this respect, he added [{\it Op. cit.}, p. 7] that, ``it would be a misconception to believe that the difficulties of the atomic theory may be evaded by eventually replacing the concepts of classical physics by new conceptual forms.'' 

\begin{enumerate}
{\bf \item[Dogma II.] Classical Representation of Physics:} The principle that one needs to necessarily presuppose the classical representation of physics in order to interpret QM and the phenomena it talks about.
\end{enumerate}

\noindent Both principles go clearly against a non-classical conceptual understanding of QM. Bohr's insistence in the necessity of only using classical physical notions can be understood in relation to his interpretation of QM as a ``rational generalization of classical mechanics'' \cite{BokulichBokulich}. 

A direct consequence of such commitment against new non-classical notions has been his  analysis of the double-slit experiment in terms of the classical notions of `wave' and `particle'. However, today we know that a quantum state $\Psi$ cannot be physically interpreted either in terms of `waves' or `particles' for diferent formal and empirical reasons. Let us briefly recall some of them. Firstly, $\Psi$ is a mathematical entity that lives in configuration space, not in classical 3-dimensional space ---which in turn would allow an interpretation in terms of the Newtonian physical notions of space and time. Consequently, $\Psi$ can represent neither a `particle' nor a `wave' which are physical notions that require a classical 3-dimensional space.\footnote{In fact, it was this reason which led Bohm to abandon, at least for a while, his own hidden variable version of QM.} Secondly, according to the orthodox Born interpretation of $\Psi$, the quantum wave function describes a probability distribution that is non-Kolmogorovian, and thus, cannot be interpreted in terms of ignorance about an ASA \cite{Redei12}. Thirdly, a `click' in a typical quantum experimental set up does not behave as if that which is producing the `click' is a `particle' or a `wave'. At the empirical level, Bell's inequalities have proven explicitly that `quantum clicks' cannot be represented in terms of a classical local-realistic theory ---to which `particles' and `waves' obviously pertain. Ever since the first experiment of Aspect, the weirdness of such non-classical `clicks' has been repeatedly confirmed. Fourthly, the Kochen-Specker theorem \cite{KS} proves that $\Psi$ does not possess definite valued properties independently of the context. On the contrary, `waves' and `particles' are noncontextual entities which, as described by classical physics in phase space, do possess definite valued properties. In short, regardless of the many references made to these classical concepts within the literature, it simply makes no sense to use the notions of `wave' and `particle' in order to interpret the orthodox formalism of QM. 

It is quite clear that the knowledge we have acquired of QM today is more detailed and accurate than the one Bohr and Einstein had at the beginning of the 20th century when they discussed the interpretation of QM in terms of {\it Gedeankenexperiments}. As a matter of fact, in the meantime, many of these imagined experiments have become testable! With the knowledge we have today, it would also seem wise to recall the old logical positivist lesson that the use of inappropriate notions within a language can only create pseudoproblems.

\section{The Bohrian Project Today: Bridging the \\Quantum Gap}

Following Bohr's footsteps, one of present philosophy of physics's major concerns remains to try to explain QM in terms of our ``common sense'' understanding of the world. As Mauro Dorato remarks \cite[p. 369]{Dorato15}, the project amounts to try to look for ``the best candidate to bridge the gap between the manifest and the physical image of the world emerging from quantum mechanics.'' Sellars characterizes the manifest image as ``the framework in terms of which man came to be aware of himself as man-in-the-world'',\footnote{\cite{deVries15}.} which, as remarked by de Vries [{\it Op. cit.}], means more broadly ``the framework in terms of which we ordinarily observe and explain our world.'' According to orthodoxy ---in close relation to Bohr's dogmas--- rather than to imagine new physical experiences or new ways of representing reality,  the philosopher should focus in ``bridging the gap'' between the new technical scientific developments and our common sense-manifest image of the world. 

The focus in trying to explain QM in terms of  ``classical reality'' and our ``common sense'' observability, is shared not only by most interpretations of QM, but also by most philosophical perspectives within the field ---both realists and anti-realists. The general scheme which allows to state this particular problem is grounded ---implicitly or explicitly--- in the distinction between  {\it observable} and  {\it non-observable} put forward by the logical positivists. Contrary to our definition of realism which considers representation as a construct of physical theories, both realist and anti-realist orthodox positions accept as a standpoint the idea that physical observation provides a direct access to reality {\it as it is}. This idea was already implicit in the logical positivist distinction between {\it theoretical terms} and {\it empirical terms}. But even though the philosophy of science community itself has characterized this distinction as ``naive'',\footnote{This distinction was strongly criticzed since the 60s by many, including Kuhn and Feyerabend.} the problems discussed in the literature today still presuppose implicitly such distinction. As remarked by Curd and Cover: 

\begin{quotation}
\noindent {\small``Logical positivism is dead and logical empiricism is no longer an avowed school of philosophical thought. But despite our historical and philosophical distance from logical positivism and empiricism, their influence can be felt. An important part of their legacy is observational-theoretical distinction itself, which continues to play a central role in debates about scientific realism.'' \cite[p. 1228]{PS}}
\end{quotation}

We must remark that even scientific realism accepts the empiricist standpoint of common sense observability. Indeed, as Musgrave \cite[p. 1221]{PS} makes the point: ``In traditional discussions of scientific realism, common sense realism regarding tables and chairs (or the moon) is accepted as unproblematic by both sides. Attention is focused on the difficulties of scientific realism regarding `unobservables' like electrons.'' In line with both Bohr, the logical positivist  `observable non-observable distinction' also closes the door to the development of new physical representations since it assumes that we already know what reality {\it is} in terms of the (naive) observation of tables and chairs ---also known as ``classical reality''. From this conservative perspective the task of the philosopher should be to restore a classical understanding about {\it what there is}. Cao's criticism is in this respect very strong:

\begin{quotation}
\noindent {\small``The old-fashioned (positivist or constructive empiricist) tradition to the distinction between observable and unobservable entities is obsolete. In the context of moden physics, the distinction that really matters is whether or not an entity is cognitively accessible by means of experimental equipment as well as conceptual, theoretical and mathematical apparatus. If a microscopic entity, such as a W-boson, is cognitively accessible, then it is not that different from a table or a chair. It is clear that the old constructive empiricist distinction between observables and nonobservables is simply impotent in addressing contemporary scientific endeavor, and thus carries no weight at all. If, however, some metaphysical category of microscopic entities is cognitively inaccessible in modern physics, then, no matter how basic it was in traditional metaphysics, it is irrelevant for modern metaphysics.''  \cite[pp. 64-65]{Cao03}}
\end{quotation}

\section{Representational Realism: A Pluralist-Ontict Approach}

Within philosophy of physics, realism has been characterized as a stance which assumes the existence of a reality independent of the actions of any human subject or concious being. In short, realism is commited to the belief of an independent reality. However, this account falls short when attempting to grasp the {\it praxis} of realist physicists themselves. Representational realism attempts to capture exactly this aspect; i.e., the specific way through which realist physicists produce a representational (meta-physical) account of reality. In this respect, the main presupposition of representational realism is that physical theories relate to reality, not only through their mathematical formalisms, but also through a network of physical concepts. The coherent interrelation between mathematical and conceptual structures allows physical theories to represent (in different ways) physical reality. In turn, every new representation allows the physicist to imagine and explore new physical phenomena. 

For the representational realist, the task of philosophy of physics is to produce physical representations which would allow her to grasp the features of the world beyond mathematical schemes and measurement outcomes. In order to provide such representation we must necessarily complement mathematical formalisms with networks of physical concepts. In this sense, for the representational realist it is not enough to claim that ``according to QM the structure of the world is like Hilbert space'' or  that ``quantum particles are like vectors in Hilbert space''. Mixing improperly classical notions with the quantum formalism is simply not doing the job of providing a conceptual physical representation in the sense discussed above. 

Our representational realist stance is intimately linked to Heisenberg's closed theory approach \cite{Bokulich04} according to which there is no need of a reductionistic understanding of inter-theory relation.\footnote{An important example to understand this non-reductionistic characterization is how Newtonian mechanics has to be understood in relation, for example, to relativity theory. According to Heisenberg \cite[pp. 97-98]{Heis58}: ``New phenomena that had been observed could only be understood by new concepts which were adapted to the new phenomena. [...] These new concepts again could be connected in a closed system. [...] This problem arose at once when the theory of special relativity had been discovered. The concepts of space and time belonged to both Newtonian mechanics and to the theory of relativity. But space and time in Newtonian mechanics were independent; in the theory of relativity they were connected.''} Every physical theory needs to develop its own conceptual scheme ---independent of those produced by different (closed) theories. The multiplicity of physical representations of reality does not invalidate by itself a realist stance. In fact, we have argued elsewhere that a Spinozist scheme would allow us to provide the missing {\it univocity condition} required to make sense of such {\it multilpe representations} as expressions of the same {\it one reality}.  

Our pluralist scheme evoids the reductionistic requirement imposed by orthodoxy. Theories are created discontinuously, through jumps. As remarked by Heisenberg in an interview by Thomas Kuhn \cite[p. 98]{Bokulich06}: ``The decisive step is always a rather discontinuous step. You can never hope to go by small steps nearer and nearer to the real theory; at one point you are bound to jump, you must really leave the old concepts and try something new... in any case you can't keep the old concepts.'' The only important aspect to consider a physical theory as `closed' is the internal {\it coherency} between the formal mathematical elements, the conceptual structure and the physical experience involved.\footnote{It is important to remark that the coherency to which we relate in this case is not the one discussed in ``coherence theory of truths'' in philosophy of science (see for example \cite{Young08}). While coherence theory of truths discuss about the relation between {\it propositions} within a physical theory, we discuss about the relation between {\it concepts}, {\it mathematical expressions} and {\it physical experience}.} The coherency reflects a wholeness present in every `closed theory' which Heisenberg beautifully expresses in the following manner:

\begin{quotation}
\noindent {\small``One finds [in closed theories] structures so linked and entangled with each other that it is really impossible to make further changes at any point without calling all the connections into question [...] We are reminded here of the artistic ribbon decorations of an Arab mosque, in which so many symmetries are realized all at once that it would be impossible to alter a single leaf without crucially disturbing the connection of the whole.'' \cite[p. 95]{Bokulich06}} \end{quotation}

Physical representation allows us to think about experience and predict phenomena without the need of actually performing any measurement. It allows us to imagine physical reality beyond the here and now.\footnote{This is of course the opposite standpoint from empiricists who argue instead that the fundament of physics is `actual experimental data'. As remarked by van Fraassen \cite[pp. 202-203]{VF80}: ``To develop an empiricist account of science is to depict it as involving a search for truth only about the empirical world, about what is actual and observable.''} Following both Einstein and Heisenberg, representational realism assumes a radical metaphysical stance regarding the possibility of observability in physics. Every new physical theory creates a new experience and thus, also new possibilities of observability. Observability is not `a given', observability in physics is always constructed, always grounded on metaphysical principles. Classical physical observation ---restricted by the (metaphysical) PE, PNC and PI (see section 1)--- is just a particular way of observing the world. A particular viewpoint grounded on classical physics and metaphysics.

According to our stance, physical theories contain a conceptual level from which physical notions allow us to {\it represent} what the theory is talking about. This basic assumption implies that mathematical structures are not enough to produce by themselves a meaningful physical representation of reality. Physics cannot be exclusively reduced to mathematical structures which predict measurement ouctomes. As Heisenberg makes the point: ``The history of physics is not only a sequence of experimental discoveries and observations, followed by their mathematical description; {\it it is also a history of concepts.}  For an understanding of the phenomena the first condition is the introduction of adequate concepts. {\it Only with the help of correct concepts can we really know what has been observed.}''\footnote{\cite[p. 264]{Heis73} (emphasis added).} As said before, in a slogan, to explain is to represent.\\

\noindent {\it {\bf Representational Realism:} A representational realist account of a physical theory must be capable of providing a physical representation of reality in terms of a network of physical concepts which coherently relates to the mathematical formalism of the theory and allows to articulate and make predictions of definite phenomena. Observability in physics is always theoretically and metaphysically laden, and thus must be regarded as a consequence of each particular physical representation.}

\section{Formalism, Concepts and Observation}

The distance bewteen orthodoxy and our representational realist project, can be understood from the radical difference in which these general schemes place the contitutive elements of a physical theory, namely, the  mathematical formalism, the network of physical concepts and the field of phenomena it is able to describe. This distance is somewhat analogous to the distance bewteen Bohr's and Einstein's philosophy of physics. While Bohr takes as a standpoint of his analysis `classical phenomena' ---or `observational terms' in philosophy of physics--- both Einstein and the representational realist consider observations and phenomena always as metaphysically laden. As Einstein remarked in a letter to Schr\"{o}dinger in the summer of 1935: ``The problem is that physics is a kind of metaphysics; physics describes `reality'. But we do not know what `reality' is. We know it only through physical description...'' \cite[p. 1196]{PS} According to the representational realist perspective, in order to truly understand a phenomenon we need to have, beforehand, an adequate set of physical notions. Only with the help of correct concepts and metaphysical principles can we really know what has been observed. This implies that there is in principle no exclusive commitment to classical phenomena. Observation is always both theory and metaphysically laden.

Even though classical obsevation ---which is in fact constrained by the metaphysical PE, PNC and PI--- is regarded by orthodoxy as the ``self evident'' standpoint for the development and analysis of theories, it is in principle possible to imagine ---specially after the creation of non-classical logics--- that different principles could be considered in order to produce a new understanding of physical reality. It is only the theory, and the metaphysical principles in which it is grounded, which can tell you what can be observed. According to the representational realist stance, empirical adequacy is only a way to decide whether the theory is adequate or not, but the main goal of physics is to provide always new representations of experience and reality. 

Physical observability is always dependent on the physical representation provided by each closed theory. From our constructive metaphysical line of research, the task of both physicists and philosophers is to jointly construct new mathematical formalisms and networks of physical concepts which allows us to imagine new physical phenomena. These three elements interrelate in such a way that the one can help the other to correct itself, to become more adequate. It is the interrelation between the mathematical structure, the physical concepts and the phenomena which allows a theory to provide a particluar representation of physical reality. As Einstein \cite[p. 175]{Dieks88a} made the point: ``[...] it is the purpose of theoretical physics to achieve understanding of physical reality which exists independently of the observer, and for which the distinction between `direct observable' and `not directly observable' has no ontological significance''. Observability is secondary even though ``the only decisive factor for the question whether or not to accept a particular physical theory is its empirical success.'' For the representational realist, empirical adequacy is part of a verification procedure, not that which ``needs to be saved''. 

We believe our representational realist stance is close in many ways to Einstein's realism who, as noticed by Howard: ``was not the friend of any simple realism'' \cite[p. 206]{Howard93}. Einstein did not consider observation as `a given'\footnote{This can be seen from the very interesting discussion between Heisenberg and Einstein \cite[p. 66]{Heis71} were the latter explains: ``I have no wish to appear as an advocate of a naive form of realism; I know that these are very difficult questions, but then I consider Mach's concept of observation also much too naive. He pretends that we know perfectly well what the word `observe' means, and thinks this exempts him from having to discriminate between `objective' and `subjective' phenomena.''} and stressed ---like the representational realist--- the importance of the metaphysical level of physical representation of reality.\footnote{See for a detailed analysis of the importance of metaphysics for Einstein \cite{Howard93}.} We also find in Einstein a focus of attention regarding the importance of physical concepts, as well as a recognition of the threat of remaining captive of ``common sense realism'':  

\begin{quotation}
\noindent {\small ``Concepts that have proven useful in ordering things easily achieve such an authority over us that we forget their earthly origins and accept them as unalterable givens. Thus they come to be stamped as `necessities of thought,' `a priori givens,' etc. The path of scientific advance is often made impossible for a long time  through such errors. For that reason, it is by no means an idle game if we become practiced in analyzing the long common place concepts and exhibiting those circumstances upon which their justification and usefulness depend, how they have grown up, individually, out of the givens of experience. By this means, their all-too-great authority will be broken. They will be removed if they cannot be properly legitimated, corrected if their correlation with given things be far too superfluous, replaced by others if a new system can be established that we prefer for whatever reason.'' \cite{Howard10}}
\end{quotation}

The representational realist stance argues in favour of a conceptual-metaphysical understanding of physical theories. Indeed, to provide a coherent metaphysical picture is the only way to truly understand what a physical theory is talking about. As Cao makes the point: 

\begin{quotation}
\noindent {\small``An important point here is that metaphysics, as reflections on physics rather than as a prescription for physics, cannot be detached from physics. It can help us to make physics intelligible by providing well-entrenched categories distilled from everyday life. But with the advancement of physics, it has to move forward and revise itself for new situations: old categories have to be discarded or transformed, new categories have to be introduced to accommodate new facts and new situations.'' \cite[p. 65]{Cao03}}
\end{quotation}

Going back to QM, since the standpoint of the representational realist stance is different from many empiricist approaches, it also confronts a different set of problems an limiting conditions. In the following section we attempt to discuss two new problems, introduced by the representational realist project, which consider explictly the representation of QM beyond mathematical structures and the prediction of measurement outcomes.

\section{The New Representational Problems of QM}

The main idea behind the representational realist stance is that when a mathematical formalism is coherently related to a network of physical concepts, it is possible to produce a physical representation of reality which allows us to describe particular physical phenomena. It is not enough to say that QM cannot be explained in terms of ``classical reality''. Neither is enough to argue that the world can be understood as a mathematical structure. This is not doing the job of producing a conceptual representation of QM. QM has proven already to be empirically adequate and the orthodox formalism is mathematically rigorous. However, instead of changing the formalism or adding {\it ad hoc} rules in order to restore a classical discourse and representation, there is a different strategy that could be considered. According to this new strategy what needs to be done is to construct a new net of (non-classical) physical concepts capable of interpreting the formalism {\it as it is}. There are some few attempts in this direction which we mentioned above. 

Contrary to our stance, the orthodox line of research deals with a specific set of problems which analyze QM from a classical perspective. This means that all problems assume as a standpoint ``classical reality'' and only reflect about the formalism in ``negative terms''; that is, in terms of the failure of QM to account for the classical representation of reality and the use of its concepts: separability, space, time, locality, individuality, identity, actuality, etc. The ``negative'' problems are thus: {\it non-}separability, {\it non-}locality, {\it non-}individuality, {\it non-}identity, etc.\footnote{I am thankful to Bob Coecke for this linguistic insight. Cagliari, July 2014.} These ``no-problems'' begin their analysis considering the notions of classical physics, assuming implicitly as a standpoint the strong metaphysical presupposition that QM should be able to represent physical reality in terms of such classical notions.

Two of the main problems discussed in the literature which attempt to provide an answer to the emergence of our manifest image from the quantum formalism are the famous measurement and basis problems. While the basis problem attempts to justify the choice of a particular basis among the many possible incompatible ones escaping the contextual character of QM, the measurement problem attempts to justify actual outcomes arising from strange superposed states. Contrary to orthodoxy, the representational realist project attempts to consider the formal features behind these problems, namely, quantum contextuality and the superposition principle, as positive elements which must guide us in the creation of a conceptual representation of QM. The replacing of these main problems and their consequences for the interpretation of QM will be discussed in more detail in the subsequent subsections.

\subsection{The Superposition Problem: Representation Beyond Actual Outcomes}

Maybe the most famous of all interpretational problems of QM is the so called ``measurement problem''. Let us briefly recall it.\\

\noindent {\it {\bf Measurement Problem:} Given a specific basis (or context), QM describes mathematically a quantum state in terms of a superposition of, in general, multiple states. Since the evolution described by QM allows us to predict that the quantum system will get entangled with the apparatus and thus its pointer positions will also become a superposition,\footnote{Given a quantum system represented by a superposition of more than one term, $\sum c_i | \alpha_i \rangle$, when in contact with an apparatus ready to measure, $|R_0 \rangle$, QM predicts that system and apparatus will become ``entangled'' in such a way that the final `system + apparatus' will be described by  $\sum c_i | \alpha_i \rangle  |R_i \rangle$. Thus, as a consequence of the quantum evolution, the pointers have also become ---like the original quantum system--- a superposition of pointers $\sum c_i |R_i \rangle$. This is why the {\it MP} can be stated as a problem only in the case the original quantum state is described by a superposition of more than one term.} the question is why do we observe a single outcome instead of a superposition of them?}\\

\noindent The measurement problem is also a way of discussing the quantum formalism in ``negative terms''. In this case, the problem concentrates in the justification of actual measurement outcomes. It should be remarked that the measurement problem presupposes that the basis (or context) ---directly related to a measurement set up--- has been already determined (or fixed). Thus it should be clear that there is no question regarding the contextual character of the theory within this specific problem. As we have argued extensively in \cite{deRonde15c}, the measurement problem has nothing to do with contextuality. The problem araises when the actualization process is considered. There is what could be described as a mix of subjective and objective elements when the recording of the experiment takes place ---as Wigner clearly exposed with his famous friend \cite[pp. 324-341]{WZ}. The problem here is the shift from the physical representation provided when the measurement was not yet performed (and the system is described in terms of a quantum superposition), to when we claim that ``we have observed a single measurement outcome'' which is not described by the theory. Since there is no physical representation of ``the collapse'', the subject (or his friend) seems to define it explicitly. The mixture of objective and subjective is due to an incomplete description of the state of affairs within the quantum measurement process (or ``collapse''). 

The focus of the measurement problem is on the actual realm of experience. In this sense, the measurement problem is an empiricist problem which presupposes the controversial idea that actual observation is perfectly well defined. However, from a representational realist stance things must be analyzed in a radically different perspective. Indeed, for the represenational realist it is only the theory which can tell you what can be observed. This means that if we are willing to truly investigate the physical representation of quantum superpositions then we will need to ``invert'' the measurement problem and focus on the formal-conceptual level ---instead of trying to justify what we observe in classical terms. Thus, attention should be focused on the physical representation of the mathematical expression instead of attempting to somehow ``save'' the measurement outcomes, justifying through {\it ad hoc} rules the ``collapse'' of the quantum superposition to one of its terms. 

The new technological era we are witnessing through quantum information processing requires that we, philosophers of QM, pay attention to the developments that are taking place today. We believe that an important help could be provided by philosophers of physics, in case we were able to develop a conceptual representation of quantum superpositions (see also \cite{deRonde16}).\\

\noindent {\it {\bf Superposition Problem:} Given a situation in which there is a quantum superposition of more than one term, $\sum c_i \ | \alpha_i \rangle$, and given the fact that each one of the terms relates through the Born rule to a meaningful physical statement, the problem is how do we conceptually represent this mathematical expression? Which is the physical notion that relates to each one of the terms in a quantum superposition?}\\
 
\noindent The superposition problem opens the possibility to truly discuss a physical representation of reality which goes beyond the classical representation of physics. Instead of keep trying ---as we have done for almost a century--- to restore dogmatically our manifest image of the world, this new problem allows us to reflect about possible non-classical solutions to the problem of interpretation of QM.

\subsection{The Contextuality Problem: Representation Beyond Classical Contexts}

Following the analysis of the measrement problem as a negative one, the basis problem is also interesting to analyse in these terms. In fact, the basis problem is also a way of discussing quantum contextuality in ``negative terms''. The problem already sets the solution through the specificity of its questioning. It presupposes that {\it there exists a path} from the ``weird'' contextual quantum formalism to a classical non-contextual experimental set up in which classical discourse holds.\\

\noindent {\it {\bf Basis Problem:} Given the fact that $\Psi$ can be expressed by multiple incompatible bases (given by the choice of a Complete Set of Commuting Observables) and that due to the KS theorem the observables arising from such bases cannot be interpreted as simultaneously preexistent, the question is: how does Nature make a choice between the different bases? Which is the objective physical process that leads to a particular basis instead of a different one?}\\

\noindent If one could explain that path through an objective physical process, then the choice of the experimenter could be regarded as well, as being part of such an objective process ---and not one that determines reality explicitly. Unfortunately, the problem remains with no solution within the limits of the orthodox formalism. Today, many interpretations attempt to ``explain'' the process with the addition of strange {\it ad hoc} rules, unjustified mathematical jumps and the like. These rules ``added by hand'', not only lack any physical justification but, more importantly, also limit the counterfactual discourse of the meaningful physical statements provided by the theory.

From the representational realist viewpoint, instead of trying to escape the contextual character of the theory, quantum contextuality should be understood as a consequence of the mathematical structure of the theory ---a formal scheme which has been allowing us for more than one century to produce the most outstanding physical predictions. The feature of contextuality emerges from the orthodox formalism of QM itself, it is not something external to it. But instead of regarding quantum contextuality as a new interesting feature of a theory, many orthodox interpretations discussed in the literature focus their efforts in trying to escape or bypass quantum contextuality so as to restore a classical noncontextual representation of physical reality. We believe that to deny contextuality just because it obstructs an interpretation of the theory in terms of actual (definite valued) properties would be tantamount to trying to deny the Lorentz transformations in special relativity simply because of its implications to the contraction of rigid rods. Indeed, this was the attempt of most conservative physicists until Einstein made the strong interpretational move of taking seriously the formalism of the theory and its phenomena, and derived a new net of physical notions in order to coherently understand the new theory. 

Instead of trying to escape quantum contextuality our representational realist stance proposes to discuss the following contextuality problem:\\

\noindent {\it {\bf Contextuality Problem:} Given the fact that Hilbert space QM is a contextual theory, the question is: which are the concepts that would allow us to coherently interpret the formalism and provide a representation of physical reality that accounts for this main feature of the theory?}\\
 
\noindent  If we accept the orthodox formalism, then contextuality is the crux of QM. It is contextuality that which needs to be physically interpreted, instead of something that needs to be bypassed because of its non-classical consequences. Like the superposition problem, the contextuality problem opens the possibility to truly discuss a physical representation of reality which goes beyond the classical representation of physics in terms of an {\it ASA}. 

While the measurement and basis problems attempt to justify ``classical reality'', independently of the quantum formalism, the superposition and contextuality leave open the possibility to the development of new non-classical schemes. Orthodox problems escape right from the start the possibility to analyse and discuss non-classical representations. This is the reason why, without a replacement of the problems addressed in the literature there is no true possibility of discussing an interpretation of QM which provides an objective non-classical physical representation of reality. We know of no reasons to believe that this is not doable. 

The failure of orthodoxy to solve the measurement and basis problems is also the failure to solve the orthodox (reductionistic) problem which attempts to bridge the gap between our common sense-classical description of reality and the quantum formalism. We will now turn our attention to this more general problem.

\section{The Quantum to Classical Limit Revisited}	

From an orthodox perspective, the most important problem that needs to be solved still remains ---after almost one century of attempts--- to explain the inter-theoretic relationship between QM and classical physics. Indeed, reductionistic approaches must explain necessarily the emergence of the classical world from QM. As John Hawthorne remarks:

\begin{quotation}
\noindent {\small ``[A] natural question to ask is how the familiar truths about the macroscopic world that we know and love (`the manifest image') emerge from the ground floor described by the fundamental book of the world. Assuming that we don't wish to concede that most of our ordinary beliefs about the physical world are false, we seem obliged to make the emergence of the familiar world from the ground floor intelligible to ourselves.'' \cite[p. 144]{Hawthorne09}} \end{quotation}

\noindent In this respect, it becomes of outmost importance to recognize the fact that the problem of `the quantum to classical limit' is a conceptual representational realist problem. It is obviously a problem which goes beyond our observable classical world. It attempts to provide a physical explanation and representation of what {\it is} the relation between the quantum and the classical realms. 

The quantum to classical limit cannot be understood from an exclusive epistemic perspective, for if there is no reference of the theory to ``something happening within physical reality'', beyond our concious recognition of measurement outcomes, the question cannot be even posed. In fact, in the explanation of the quantum to classical limit the subject is simply out of the picture. If we assumed, from an epistemic viewpoint, that physical theories provide ``an economical account of experience'' with no metaphysical referent whatsoever, then there seems to be no interesting `limit' or `relation' to be discussed between these two theories. The problem simply disapears. As a matter of fact, both classical mechanics and QM (within their specific limit of applicability) already accomplish their means with respect to their empirical findings. Let us be explicit about this point.  The problem of the quantum limit addresses the description and representation of a real physical process, it seeks to explain the conceptual representation of what is exactly going on within physical reality. Any realist description must be provided necessarily beyond mathematical structures and measurement outcomes, for if there is no conceptual metaphysical level of analysis the ``solution'' seems to differ in no way from instrumentalism itself.   

The reductionistic perspective requires an explanation of the limit from QM to ``classical reality''. Whithout a clear explanation of such limit orthodoxy seems to run into trouble. The limit cannot be regarded exclusively in terms of a formal or empirical reduction. It is also a problem which requires necessarily a conceptual level of analysis capable of providing a physical explanation of such a seemingly incompatible relation. And the explanation of a physical relation between theories seems to necessarily imply the understanding of the relata. In order to explain the path from QM to classical mechanics it seems one should understand, first of all, what is QM really talking about. 

Bohr himself made especial emphasis on the fact that the measurement process was the key to recover a ``rational account of physical phenomena'' \cite{BokulichBokulich}. In line with Bohr's concerns, the principle of decoherence introduced by Zurek in 1981, attempted to provide the missing physical explanation of the ``the quantum to classical limit'' \cite{Zurek81, Zurek82}. Indeed, would there be an objective explanation of such physical process that would turn ``quantum particles'' into ``stable macroscopic objects'', it would also imply a solution to the basis and measurement problems of QM. Such objective explanation would then provide the key to a unified (reductionistic) representation of physical reality which would contain both quantum and classical physics.

As remarked by Jeffrey Bub, in the last decades decoherence has become ``the new orthodoxy'' \cite[p. 212]{Bub97}. The popularity of the principle came from the repeated claims which argued that decoherence provided a suitable solution to the problem of the quantum to classical limit. As exaplined by Zurek \cite[p. 20]{Zurek02}: ``[Classical] reality emerges from the substrate of quantum physics: Open quantum systems are forced into states described by localized wave packets. They obey classical equations of motion, although with damping terms and fluctuations that have a quantum origin. What else is there to explain?'' However, when decoherence theorists reflect about the physical meaning of such isolated quantum states, things become quite bizarre:

\begin{quotation}
\noindent {\small ``If the unknown state cannot be found out ---as is indeed the case for isolated quantum systems--- then one can make a persuasive case that such states are subjective, and that quantum state vectors are merely records of the observer's knowledge about the state of a fragment of the Universe (Fuchs and Peres 2000). However, einselection is capable of converting such malleable and `unreal' quantum states into solid elements of reality.'' 
[{\it Op. cit.}, p. 22]} \end{quotation}

\noindent Indeed, it is claimed by Zurek that decoherence is capable of creating the ``real'' from the ``unreal'', objective states from subjective choices. 

\begin{quotation}
\noindent {\small``Quantum state vectors can be real, but only when the superposition principle ---a cornerstone of quantum behavior--- is `turned off' by einselection. Yet einselection is caused by the transfer of information about selected observables. Hence, {\it the ontological features of the state vectors ---objective existence of the einselected states--- is acquired through the epistemological `information transfer'.}'' [{\it Op. cit.}, p. 22] (emphasis added)} \end{quotation}
 
Unfortunatelly, regardless of Zurek's strong claims, it was soon recognized that the promise of the new orthodoxy to account for the quantum to classical limit was neither physically nor philosophically justified. In fact, the principle of decoherence has been found to have many problems, {\it ad hoc} moves as well as unjustified shortcuts. There are several important technical problems which cast severe doubts on the validity of the original project of decoherence to explain the limit. Let us mention some of them: 

\begin{enumerate}
\item{I.} The fact that the diagonalization of quantum states is not complete, since ``very small'' is obviously not ``equal to zero''. It is true that decoherence seems to produce approximately diagonal mixtures in very ``short times'', and it is also true that from an epistemic viewpoint ``very small'' might be considered as superfluous when compared to ``very big''; however, this is clearly not  the case when considering an ontological problem such as the one addressed by the limit. From an ontological perspective there is no essential difference between ``very big'' and ``very small'', they both have exactly the same importance. The problem is not if a subject is capable of observing something ``big'' or ``small'', the problem is to provide a coherent representation ---both formal and conceptual--- of {\it what there is} ---even if it is very, very small! 

\item{II.} The fact that the diagonalization can recompose itself into un-diagonalized mixtures if enough time is considered \cite{Recoherence, CormikPaz08}. The restitution of non-diagonal terms comprises the whole program since the fact that time is ``long'' or ``short'' is neither important from an ontological perspective. Once again, the problem is not epistemic, it is not about {\it what a subject might be able to observe}; the problem of the limit is ontic problem, it is a question about {\it what there is in physical reality} ---independently of conscious beings. 

\item{III.} The fact that the principle only allows to turn ---at least for a while--- non-diagonal improper mixtures into ``approximately'' diagonal improper mixtures does not solve anything. Even if decoherence was able to accomplish the complete diagonalization ---something it does not--- of the improper mixture, the problem remains that all improper mixtures simply cannot be interpreted in terms of ignorance. As remarked by Zhe:

\begin{quotation}
\noindent {\small``[...] by decoherence I mean the practically irreversible and
practically unavoidable (in general approximate) disappearance of certain phase relations from the states of local systems by interaction with their environment according to the Schr\"odinger equation. Since phase relations cannot absolutely disappear in a unitary evolution, this disappearance can only represent a delocalization, which means that the phases ``are not there'' any more, neither in the system nor in the environment, although they still exist in the total state that describes both of them in accordance with quantum nonlocality [...]  Decoherence is thus a normal consequence of interacting quantum mechanical systems. It can hardly be denied to occur ---but it cannot explain anything that could not have been explained before.'' \cite{Zhe96} } \end{quotation}

\item{IV.} The late recognition by Zurek of the fact that the principle of decoherence only leads to (approximately diagonal) improper mixtures has forced him to venture himself into Everett's many worlds interpretation and quantum bayesianism (see \cite{Zurek02}). However, also here, as it has been discussed by Dawin and Th\'ebault, serious inconsistencies threaten the project \cite{DawinThebault15}. 

\item{V.} Even more worrying is the analysis provided by Kastner, who has pointed out that ---even if all these just mentioned problems would be left aside--- the main reasoning of the whole decoherence program is circular \cite{Kastner}. There has been, up to date, no reply to the analysis provided by Kastner.  
\end{enumerate}

Due to all these well knwon facts about decoherence, today it is becoming to be more and more recognized, mainly within the specialized literature, that the principle of decoherence has failed to provide a convincing physical explanation of the quantum to classical limit (see \cite{Joos}).  As remarked by Guido Bacciagaluppi: 

\begin{quotation}
\noindent {\small``[some physicists and philosophers] still believe decoherence would provide a solution to the measurement problem of quantum mechanics. As pointed out by many authors, however (e.g. Adler 2003; Zeh 1995, pp. 14-15), this claim is not tenable. [...] Unfortunately, naive claims of the kind that decoherence gives a complete answer to the measurement problem are still somewhat part of the `folklore' of decoherence, and deservedly attract the wrath of physicists (e.g. Pearle 1997) and philosophers (e.g. Bub 1997, Chap. 8) alike.'' \cite{Bacciagaluppi12}} \end{quotation}

In fact, today, there seems to be more questions than answers when considering the solution provided by decoherence to the quantum to classical limit. So it seems, the principle of decoherence might be regarded, in this specifc respect, at best, as a proto-principle or proto-model of the quantum process of measurement, but never ---at least at this stage of its development--- as a coherent physical representation of the path from the quantum to the classical. This is not to undermine the importance of decoherence. One might recall that proto-models have been of great importance in the development of physical theories. Even in QM, we find a very good example of the importance of proto-models when recalling the history behind Bohr's atomic model. Quite regardless of its empirical success this model was abandoned due to the incoherent physical descrpition it provided. The critical recognition by Heisenberg and Pauli of the failure of Bohr's atomic model to produce a coherent formal and conceptual account might be regarded as the very precondition of possibility for the development of QM itself. We could say that matrix mechanics and Schr\"odinger's formulation were developed because it was recognized that the Bohrian model was simply not a closed physical theory. 

\section{Is the `FAPP Solution' a `Solution'?} 

Unfortunatelly, after more or less having recognized that decoherence does not solve that which it had promised to solve originally, something very strange happened within the new community. Instead of reconsidering the problem and the set of presuppositions involved in order to re-develop the principle of decoherence, a new justification was advanced by the new orthodoxy. Even though it was accepted that decoherence did not ``really'' solve the quantum to classical limit, it was argued that ``the principle of decoherence solves the problem For All Practical Purposes (FAPP)!'' This has been a way of the new orthodoxy to argue ---more or less in disguise--- that ``we cannot really explain the path in physical terms, but don't worry, it works anyhow!'' This instrumentalist justification escapes any physical explanation grounding itself, once again, in the predictive power of the theory ---a predictive power no physicist had ever doubted in the first place. 

Following an epistemic viewpoint it might then be argued that decoherence is an ``economy of experience'', that it helps us to work in the lab, that it is in fact an ``epistemic solution''. But then the original ontological problem has been completely abandoned, the original question completely forgotten. This new ``FAPP solution'' ---which is in itself a revival of instrumentalism placed in the heart of realist discussions--- has been imported in order to discuss the ontological question of the quantum to classical limit, not only without a proper recognition of the failure of the original project but also turning the original problem into a pseudoproblem. This re-cooked quantum omelette produced by decoherence ---which like many interpretations today create physical reality from epistemic choices--- has penetrated physics so deeply that today many physicist seem to uncritically accept there actually exists a physical process called ``decoherence'' that {\it really} takes place in the lab. But when asked to explain what is exactly this physical process about, the new orthodoxy rapidly shifts the debate and using this new instrumentalist justification argue: ``decoherence works FAPP!''   

The main problem of the `FAPP solution' is that it solves nothing, it just sweeps the (quantum) dirt under the (classical) carpet. By repeating that ``it works!'', many physicist and philosophers believe the problem has been actually solved. And there is nothing less interesting in physics than engaging in a problem that already has a solution. There is nothing there to be done, nothing to be thought or in need of development. But problems are the true gas of science, they are that which scientist work on, concentrate with passion, allow us to produce new physical theories. Problems in science should not to be regarded as ghosts or monsters that we need to destroy. In fact, there is nothing more interesting, more encouraging than a good difficult scientific problem. 

In QM, it was Heisenberg and Pauli's unsatisfaction with Bohr's ``magical'' model of the atom ---as Sommerfeld used to call it--- which led them to develop matrix mechanics and the exclusion principle. Instead of confronting the problem, orthodoxy has advanced the most weird type of justifications: ``more or less solved'', ``approximately solved'', ``almost solved'', ``solved FAPP''? These solutions simply escape the question at stake. But in fact, it is quite simple, ontological problems canot be solved epistemically. A representational realist problem, such as the quantum to classical limit, cannot be solved FAPP.

\section*{Acknowledgements} 

I want to thank private discussions with R. Kastner and G. Domenech regarding these subjects. This work was partially supported by the following grants: FWO project G.0405.08 and FWO-research community W0.030.06. CONICET RES. 4541-12.


\end{document}